\documentstyle[epsfig]{aipproc}

\begin{document}
\title{Event Rates for Binary Inspiral}
\author{Vassiliki Kalogera$^*$ and Krzysztof Belczynski$^{*,\dagger}$}
\address{$^*$Harvard-Smithsonian Center for Astrophysics, Cambridge, MA
02138\\
$^{\dagger}$Nicolaus Copernicus Astronomical Center, 00-716 Warszawa,
Poland}
\maketitle

\begin{abstract}

Double compact objects (neutron stars and black holes) found in binaries
with small orbital separations are known to spiral in and are expected to
coalesce eventually because of the emission of gravitational waves.  Such
inspiral and merger events are thought to be primary sources for ground
based gravitational--wave interferometric detectors (such as LIGO). Here,
we present a brief review of estimates of coalescence rates and we examine
the origin and relative importance of uncertainties associated with the
rate estimates. For the case of double neutron star systems, we compare
the most recent rate estimates to upper limits derived in a number of
different ways. We also discuss the implications of the formation of
close binaries with two non--recycled pulsars. 

\end{abstract}

\section*{Introduction}

Compact objects, neutron stars (NS) or black holes (BH), formed from
relatively massive stars can spiral in and coalesce when found in tight
binaries, the
orbital evolution of which is driven by gravitational radiation. As
angular momentum losses dominate, the orbit shrinks and the two compact
objects can eventually merge as they revolve in orbit around each
other. The prototype progenitor system of such inspiral events is the
binary pulsar PSR B1913+16 (the ``Hulse--Taylor'' pulsar \cite{HT75}).
Sensitive pulsar timing measurements have revealed that the orbital period
decreases at a rate comparable (to better than 1\%) to that predicted by
general relativity for the emission of gravitational waves \cite{TW82},
\cite{TW89}. The ultimate coalescence of the two neutron stars seems
inevitable.

Although PSR B1913+16 will not reach coalescence for another 300\,Myr,
similar inspiraling systems in the Milky Way and nearby galaxies are
thought to be primary sources of gravitational radiation for ground--based
interferometric gravitational--wave detectors, currently under
construction or commissioning (e.g., LIGO, VIRGO, GEO600). In addition to
NS--NS close binaries, NS--BH and BH--BH binaries are also expected to
form through the evolution of massive binaries and to contribute to the
detection of inspiral events.

The expected detection rate of inspiral events depends on (i) the strength
of the expected gravitational--wave signal, (ii) the gravitational--wave
detector sensitivity, and (iii) the coalescence rate of each binary
population. The first two considerations define a maximum distance $D_{\rm
max}$, out to which different types of inspiral events and mergers could
be detected. The coalescence rate for each population is estimated in two
steps: first, the Galactic rate, and then its extrapolation out to the
maximum distance of interest. Based on the current understanding of the
LIGO sensitivities, the maximum distances out to which inspiral events
could be detected by LIGO II (and LIGO I) are (approximately), 350\,Mpc
(20\,Mpc) for NS--NS binaries, 700\,Mpc (40\,Mpc)  for NS--BH binaries,
and 1500\,Mpc (100\,Mpc) for BH--BH binaries (assuming 1.4\,M$_\odot$ NS
and 10\,M$_\odot$ BH; Sam Finn, private communication). Given our current
best knowledge (based on recent redshift surveys) of galaxy distributions
out to those distances \cite{K00}, it can be estimated that, for a LIGO II
detection rate of 1 event per year, the following {\em Galactic}
coalescence rates are required: $\simeq 5\times 10^{-7}$\,yr$^{-1}$ for
NS--NS binaries, $\simeq 5\times 10^{-8}$\,yr$^{-1}$ for NS--BH binaries,
and $\simeq 5\times 10^{-9}$\,yr$^{-1}$ for BH--BH binaries.

Formation rates of {\em coalescing} compact binaries (systems with tight 
enough orbits that merge within a Hubble time $\sim 10^{10}$\,yr) have
been calculated so far using two very different methods:  either entirely  
theoretically, based on binary evolution models, or, for NS--NS binaries,
empirically, based on the observed NS--NS sample. In what follows we
present an up--to--date review of current rate estimates, addressing in
detail the most important uncertainties associated with them. We also
discuss independent ways of obtaining upper limits to the coalescence rate
of NS--NS binaries and possible implications of the formation of systems
without recycled pulsars. 

\section*{Theoretical Rate Estimates}

The formation rate of coalescing binary compact objects can be calculated,
given a sequence of evolutionary stages leading to binary compact object
formation.  Over the years, a relatively standard picture has been formed
describing the birth of such systems based on considerations of NS--NS
binaries \cite{BvdH91}. More recently, variations of the standard
evolutionary channel have also been discussed and suggested \cite{B95},
mainly based on worries about the fate of neutron stars in situations of
hypercritical accretion (not limited to the photon Eddington rate), and
their possible collapse into black holes. In all versions, however, the
main picture remains the same: the initial binary progenitor consists of
two binary members massive enough to eventually collapse into a NS or a
BH. The evolutionary path involves multiple phases of stable or unstable
mass transfer, common--envelope phases (where one or possibly two stellar
cores spiral in the envelopes of evolved stars and eventually lead to the
ejection of these envelopes), and accretion onto compact objects, as well
as two core collapse events. The final outcome of interest is the
formation of binary compact objects in close binary orbits.

Such theoretical modeling has been undertaken by a number of different
groups by means of population syntheses. This provides us with {\em ab
initio} predictions of coalescence rates. Monte Carlo numerical techniques
are employed in following the evolution of a large ensemble of primordial
binaries with certain assumed initial properties through a multitude of
channels until compact object binaries are formed.  The changes in the
properties of the binaries at the end of each stage are calculated based
on our current understanding of the various evolutionary processes
involved: wind mass loss from massive hydrogen-- and helium--rich stars,
mass and angular--momentum losses during mass transfer phases, dynamically
unstable mass transfer and common--envelope evolution, effects of highly
super--Eddington accretion onto NS, and supernova explosions with kicks
imparted to newborn NS or even BH. Given our limited understanding of some
of these phases, the results of population synthesis are expected to
depend on the assumptions made in the treatment of the various processes.
Therefore, exhaustive parameter studies are required by the nature of the
problem.

Recent studies of the formation of compact objects and calculations of
their Galactic coalescence rates (\cite{LPP97}, \cite{FBB98}, \cite{PZ98},
\cite{BB98}, \cite{FWH99}, \cite{BBZ99} ) have explored the input
parameter space and the robustness of the results at different levels of
(in)completeness. Almost all of these groups have studied the sensitivity
of the predicted coalescence rates to the average magnitude of the kicks
imparted to compact objects at birth. The range of predicted NS--NS
Galactic rates obtained by varying the kick magnitude alone is found in
the range $<10^{-7}~-~5\times 10^{-4}$\,yr$^{-1}$. This large range
indicates the importance of supernovae (two in this case) in the evolution
of massive binaries. Variations in the assumed mass--ratio distribution
for the primordial binaries can {\em further} change the predicted rate by
about a factor of $10$, while assumptions of the common--envelope phase
add another factor of about $10-100$. Variation in other parameters
typically affects the results by factors of two or less. Predicted rates
for BH--NS and BH--BH binaries lie in the ranges $<
10^{-7}~-~10^{-4}$\,yr$^{-1}$ and $<10^{-7}~-~10^{-5}$\,yr$^{-1}$,
respectively when the kick magnitude to both NS and BH is varied. Other
uncertain factors such as the critical progenitor mass for NS and BH
formation lead to variations of the rates by factors of $10-50$.

It is evident that recent theoretical predictions for coalescence rates
cover a wide range of values (typically 3--4 orders of magnitude), because
the various input parameters and assumptions affect strongly the absolute
normalization (birth rate) of the modeled populations. Given these
results, it seems fair to say that, at least at present, population
synthesis calculations have a rather limited predictive power and provide
fairly loose constraints on coalescence rates. One way to improve the
reliability of such predictions is to study a number of different binary
populations (with or without compact objects) and incorporate a number of
independent observational constraints, such as star formation rate,
supernova rates of different types, binarity of Wolf--Rayet stars, and
others. A number of constraints on the population synthesis models could
help restricting the predicted coalescence rates in narrower ranges
\cite{BKB00}.

\section*{Empirical Rate Estimates} 

The large range of theoretically predicted Galactic coalescence rates of
double compact objects motivates us to examine other ways of obtaining
rate estimates.  The observed sample of coalescing NS--NS binaries found
in the Galactic field (PSR B1913+16 and PSR B1534+12) provides us with
alternative estimates of their coalescence rate. ``Empirical'' estimates
can be obtained using the observed pulsar and binary properties along with
models of selection effects in radio pulsar surveys \cite{P91},
\cite{N91}. For each observed object, a scale factor can be calculated
based on the fraction of the Galactic volume within which pulsars with
properties identical to those of the observed pulsar could be detected by
any of the radio pulsar surveys, given their detection thresholds. This
scale factor is a measure of how many more pulsars like the ones detected
in the coalescing NS--NS systems exist in our galaxy. The coalescence rate
can then be calculated based on the scale factors and estimates of
detection lifetimes summed up for all the observed systems. Based on this
method the first two studies concluded that the NS--NS Galactic
coalescence rate is $\simeq 10^{-6}$\,yr$^{-1}$.

Since then, estimates of the NS--NS coalescence rate have known a
significant downward revision primarily because of (i) the increase of the
Galactic volume covered by radio pulsar surveys with no additional
coalescing NS--NS being discovered \cite{CL95}, (ii) the increase of the
distance estimate for PSR B1534+12 based on measurements of post-Newtonian
parameters \cite{S98} (iii) revisions of the lifetime estimates
\cite{vdHL96}, \cite{A98}.  Recent estimates place the NS--NS rate for our
Galaxy in the range $\simeq 1-3 \times 10^{-7}$\,yr$^{-1}$. Further, it
has been realized that a number of upward correction factors must be
included, most importantly to account (i) for the beamed nature of pulsar
emission and correct for all the binary pulsars with beams that our line
of sight does not intersect, and (ii) for the faint end of the pulsar
luminosity function and correct for those systems that are too faint to be
detected. These two correction (multiplication) factors have so far
typically been assumed to be $\simeq 3$ and $\simeq 10$, respectively.

In a study just recently completed \cite{K00}, we especially focused on
all the uncertainties associated with these empirical estimates. We found
that the upward correction factor for the faint end of the pulsar
luminosity is the most important source of uncertainty. However, it is
highly sensitive to the number of observed objects and its distribution
function widens dramatically for small--number samples. For a sample of
two objects (as the observed one) the faint--pulsar correction factor can
vary from very small (close to unity) to as high as $\simeq 200$ (see
following subsection). Beyond the issue of faint pulsars, we considered a
number of uncertainties and correction factors. Based on recent
observational data for both PSR B1913+16 and PSR B1534+12, we found that
the beaming correction factor is higher than previously thought ($\simeq
6$) but with a rather small uncertainty ($\simeq 10$\%).  Other factors,
such as pulsar ages and lifetimes, and spatial distribution, lead to an
uncertainty factor of about 2.  We estimate the Galactic NS--NS coalescence
rate in the range $\simeq 10^{-6}-5\times 10^{-4}$\,yr$^{-1}$, which is
still narrow compared to the range covered by the theoretical estimates.

\subsection*{Small Number Sample and Pulsar Luminosity Function}

One important limitation of empirical estimates of the coalescence rates
is that they are derived based on {\em only two} observed NS--NS systems,
under the assumption that the observed sample is representative of the
true population, particularly in terms of their radio luminosity. Assuming
that the recycled pulsars in NS--NS binaries follow the radio luminosity
function of young pulsars and that therefore their true Galactic
population is dominated in number by low--luminosity pulsars, it can be
shown that the current empirical estimates most probably {\em
under}estimate the true coalescence rate.  If a small--number sample is
drawn from a parent population dominated by low--luminosity (hence hard to
detect) objects, it is statistically more probable that the sample will
actually be dominated by objects from the high--luminosity end of the
population. The result is that the population overall is thought to be
brighter than it really is, and therefore, detectable over a larger
Galactic volume. Consequently, the empirical estimates based on such a
sample will tend to overestimate the detection volume for each observed
system, and therefore underestimate the scale factors and the resulting
coalescence rate.

 \begin{figure}
 \centerline{\epsfig{file=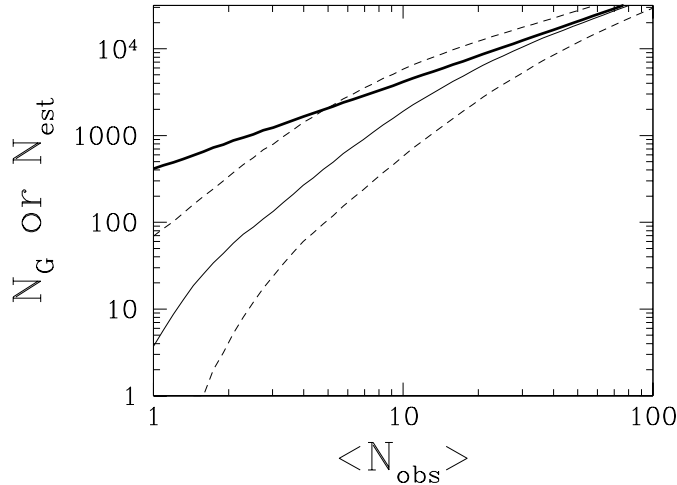,height=3.5in,angle=-90}}
 \vspace{10pt}
 \caption{Bias of the empirical estimates of the NS--NS coalescence rate
because of the small--number observed sample. See text for details.} 
 \label{fig1}
 \end{figure}

This effect can be clearly demonstrated with a Monte Carlo experiment
\cite{K00} using simple models for the pulsar luminosity function and the
survey selection effects. As a first step, the average observed number of
pulsars is calculated given a known ``true'' total number of pulsars in
the Galaxy (thick-solid line in Figure \ref{fig1}). As a second step, a
large number of sets consisting of ``observed'' (simulated) pulsars are
realized using Monte Carlo methods. These pulsars are drawn from a Poisson
distribution of a given mean number ($<N_{\rm obs}>$) and have
luminosities assigned according to the assumed luminosity function.  
Based on each of these sets, one can estimate the total number of pulsars
in the Galaxy using empirical scale factors, as is done for the real
observed sample. The many (simulated) ``observed'' samples can then be
used to obtain the distribution of the estimated total Galactic numbers
($N_{\rm est}$) of pulsars. We find that these $N_{\rm est}$ distributions
are very strongly skewed and lead to possible correction factors for the
faint pulsars in a wide range of values (covering typically a couple of
orders of magnitude). The median and 25\% and 75\% percentiles of this
distribution are plotted as a function of the assumed number of systems in
the (fake) ``observed'' samples in Figure \ref{fig1} (thin--solid and
dashed lines, respectively).

It is evident that, in the case of small--number observed samples (less
than $\sim 10$ objects), the estimated total number, and hence the
estimated coalescence rate, can be underestimated by a significant factor.
For observed samples with an expected number of objects equal to two, for
example, the true rate may be much higher by more than a factor as high as
$\simeq 200$. This underestimation factor represents an upward correction
factor that must be applied to the rate estimated using the observed
sample of {\em coalescing} NS--NS binaries. However, we note that
distribution of this correction factor covers a wide range and becomes
highly skewed for small number samples (less than about 10 objects), and
therefore it is currently quite uncertain.  We conclude that correcting
for the undetected, faint pulsars in the population cannot be decoupled
from the problems of a small--number sample because of the assumption of
the observed sample being representative of the population, implicit in
the method.

\section*{Upper Limits on the NS--NS Coalescence Rate}

 \begin{figure}
 \centerline{\epsfig{file=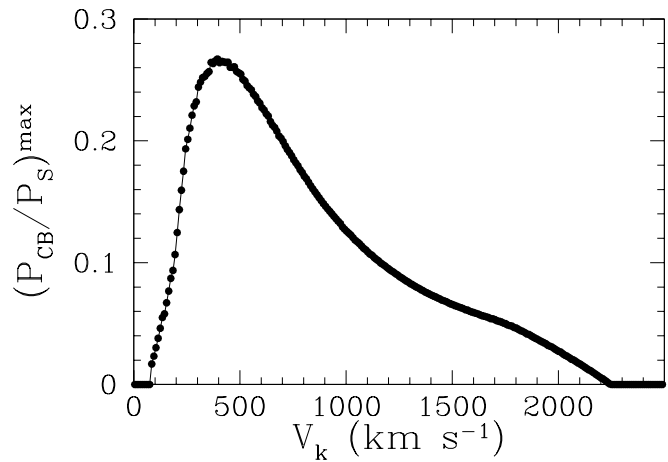,height=3.5in,angle=-90}}
 \vspace{10pt}
 \caption{Maximum probability ratio for the formation of coalescing NS--NS
systems and the disruption of binaries as a function of the kick magnitude
at the second supernova.}
 \label{fig2}
 \end{figure}

Observations of NS--NS systems and isolated pulsars related to NS--NS
formation allow us to obtain upper limits on their Galactic coalescence
rate in a number of different ways. Depending on how their value compares
to the Galactic rate required for a LIGO II detection rate of 1 event per
year, such limits can in principle provide us with valuable information
about the prospects of gravitational--wave detection.

The absence of any young pulsars detected in NS--NS systems was used to
obtain a rough upper limit to the rate of $\sim 10^{-5}$\,yr$^{-1}$
\cite{B96}. Recently the same basic argument was reexamined in more detail
and a more robust upper limit of $\sim 10^{-4}$\,yr$^{-1}$ was derived
\cite{A98}.

An upper bound to the NS--NS coalescence rate can also be obtained by
combining our theoretical understanding of orbital dynamics (for
supernovae with NS kicks in binaries)  with empirical estimates of the
birth rates of {\em other} types of pulsars related to NS--NS formation
\cite{KL00}. Progenitors of NS--NS systems experience two
supernova explosions. The second supernova explosion (forming the NS that
is {\em not} observed as a pulsar) provides a unique tool for the study of
NS--NS formation, since the post--supernova evolution of the system is
simple, driven only by gravitational--wave radiation. There are three
possible outcomes after the second supernova: (i) a coalescing NS--NS is
formed (CB), (ii) a wide NS--NS (with a coalescence time longer than the
Hubble time) is formed (WB), or (iii) the binary is disrupted (D) and a
single pulsar similar to the ones seen in NS--NS systems is ejected. Based
on supernova orbital dynamics we can accurately calculate the probability
branching ratios for these three outcomes, $P_{\rm CB}$, $P_{\rm WB}$, and
$P_{\rm D}$. For a given kick magnitude, we can calculate the maximum
ratio $(P_{\rm CB}/P_{\rm D})^{\rm max}$ for the complete range of
pre-supernova parameters defined by the necessary constraint $P_{\rm
CB}\neq 0$ (Figure \ref{fig2}). Given that the two types of systems have a
common parent progenitor population, the ratio of probabilities is equal
to the ratio of the birth rates $(BR_{\rm CB}/BR_{\rm D})$.

We can then use (i) the absolute maximum of the probability ratio ($\simeq
0.27$ from Figure \ref{fig2}) and (ii) an empirical estimate of the birth
rate of single pulsars similar to those in NS--NS based on the current
observed sample to obtain an upper limit to the coalescence rate. The
selection of this sample involves some subtleties \cite{KL00}, and the
analysis results in $BR_{\rm CB} < 1.5\times 10^{-5}$\,yr$^{-1}$. Note
that this number could be increased because of the small--number sample
and luminosity bias, which this time affects the empirical estimate of
$BR_{\rm D}$ by a factor of $\simeq 2-6$. Such an upward correction can
bring the upper limit in the range $3-9 \times 10^{-5}$\,yr$^{-1}$.

This is an example of how we can use observed systems other than NS--NS to
improve our understanding of their coalescence rate. A similar calculation
can be done using the wide NS--NS systems instead of the single pulsars
\cite{KL00}.

\section*{Non--recycled double neutron stars}

We have already pointed out that the empirical methods employed to obtain
rate estimates for NS--NS coalescence include the implicit assumption that
the properties of the observed sample are representative of the Galactic
NS--NS population. This assumption extends to the pulsar properties and
their evolutionary history of recycling (spin--up by accretion).
Consistent with the pulsars observed in the detected NS--NS systems, it
turns out that so far theoretical studies of NS--NS formation have
considered systems where one of the neutron stars had the opportunity to
be recycled, at least in principle (through stellar winds, Roche--lobe
overflow accretion, or even possibly in a common--envelope phase). 

Here, we report on a new evolutionary path leading to the formation of
close NS--NS binaries, with the unique characteristic that none of the two
NS ever had the chance to be recycled by accretion. As we will discuss in
more detail, such NS--NS systems have a negligible probability of being
detected as binary pulsars, and could represent a ``dormant'' NS--NS
population in galaxies with important implications for gravitational--wave
detection of NS--NS inspiral events. The existence of this recently
identified \cite{BK00} evolutionary channel stems from the evolution of
helium--rich stars (cores of massive NS progenitors), which has been
neglected in most previous studies of double compact object formation. We
find that these non--recycled NS--NS binaries are formed from bare
carbon--oxygen cores in tight orbits, with formation rates comparable to
or maybe even higher than those of recycled NS--NS binaries.

\subsection*{The Method} 

We study NS--NS binaries formed through a multitude of evolutionary
sequences that are not predefined, but instead are realized in Monte Carlo
population synthesis calculations. 

To describe the evolution of single stars (hydrogen-- and helium--rich) 
from the zero age main sequence (ZAMS) to carbon--oxygen (CO) core formation, 
we employ analytical formulae from stellar evolution fits \cite{H00} 
However, we have adopted a prescription for the masses of compact objects 
formed at core--collapse events, based on the relation between CO core 
masses and final FeNi core masses \cite{W86}.  

Concerning the evolution of interacting binaries, we model the changes of
mass and orbital parameters taking into
account mass and angular momentum transfer between the stars or loss from
the system during Roche--lobe overflow, tidal circularization,
rejuvenation of stars due to mass accretion, wind mass loss from massive
and/or evolved stars, dynamically unstable mass transfer episodes leading
to common--envelope (CE) evolution and spiral--in of the stars. 
We also account for the {\em possibility} of hyper-critical accretion onto
compact objects  during CE phases \cite{B95} and effects of 
asymmetric supernovae (SN) on a binary orbit (mass loss and a kick velocity 
a newly born compact object receives in SN).
More details about the treatment of various evolutionary processes are 
presented elsewhere \cite{BKB00}.

In the synthesis calculations, we typically evolve a few million of
primordial binaries to satisfy the requirement that the statistical
(Poisson) fractional errors ($\propto 1/\sqrt N$) of the final NS--NS
population are lower than 10\%.  The formation rates are calibrated using
the latest Type II SN empirical rates and normalized to our Galaxy
\cite{CET99}.

In our standard model, primordial binaries follow given distributions: 
for primary masses ($5-100$\,M$_\odot$), $\propto M_1^{-2.7}dM_1$; 
for mass ratios ($0<q<1$), $\propto dq$; 
for orbital separations (from a minimum, so both ZAMS stars fit within
their Roche lobes, up to $10^5$\,R$_\odot$), $\propto dA/A$; 
for eccentricities, $\propto 2e$. 
Each of the models is also characterized by a set of assumptions, which, 
for our standard model, are:
 (1) {\em Kick velocities.} We use a weighted sum of two Maxwellian
distributions with $\sigma=175$\,km\,s$^{-1}$ (80\%) and
$\sigma=700$\,km\,s$^{-1}$ (20\%) \cite{CC97};
 (2) {\em Maximum NS mass.} We adopt a conservative value of $M_{\rm
max}=3$\,M$_\odot$ \cite{KB96}; 
 (3) {\em Common envelope efficiency.} We assume $\alpha_{\rm
CE}\times\lambda = 1.0$, where $\alpha$ is the efficiency with which
orbital energy is used to unbind the stellar envelope, and $\lambda$ is a
measure of the central concentration of the giant;
 (4) {\em Non--conservative mass transfer.} In cases of dynamically stable
mass transfer between non--degenerate stars, we allow for mass and angular
momentum loss from the binary \cite{P92},
assuming that half of the mass lost from the donor is also lost from the
system ($1-f_{\rm a}=0.5$) with specific angular momentum equal to
$\beta2\pi$A$^2$/P ($\beta=1$);
 (5) {\em Star formation history.} We assume that star formation has been
continuous in the disk of our Galaxy for the last 10\,Gyr \cite{G01}.

\subsection*{Results}

We use our population synthesis models to investigate all possible
formation channels of NS--NS binaries realized in the simulations. We find
that a significant fraction of {\em coalescing} NS--NS systems are formed
through a new, previously not identified evolutionary path.
The evolution along this new channel begins with two phases of
Roche--lobe overflow. The first, from the primary to the secondary,
involves non--conservative but dynamically stable mass transfer
and ends when the hydrogen envelope is consumed.  The second, from the
initial secondary to the helium core of the initial primary, involves
dynamically unstable mass transfer, i.e., CE evolution. 
The post--CE binary consists of two bare helium stars of relatively low
masses. As they evolve through core and shell helium burning, the two
stars develop `giant--like'' structures, with clear CO cores and
convective envelopes. Their radial expansion eventually brings them into
contact and the system evolves through a double CE phase 
(similar to Brown [1995], for hydrogen--rich stars). During this double CE
phase, the combined helium envelopes are ejected at the expense of orbital
energy. The tight, post--CE system consists of two CO cores, which
eventually end their lives as Type Ic supernovae leaving double neutron star
system. 

The unique qualitative characteristic of this NS--NS formation path is
that both NS have avoided recycling. 
Based on comparison of non--recycled NS--NS {\em relative} to 
that of recycled pulsars, for each of our models, we derive a correction 
factor for empirical estimates of the Galactic NS--NS coalescence rate.  
Since these estimates account only for NS--NS systems with recycled pulsars, 
they must be increased to include any non--recycled systems formed.
We have performed an extensive parameter study to assure robustness of our
results.
In Table 1 we present the formation rates of non--recycled NS--NS binaries
and the total NS--NS population with merger times shorter than 10\,Gyr,
along with the upwards correction factor for the Galactic empirical rate
estimates. Results are shown only for models where the derived factor
differs from our standard model by more than 25\%. 
We find that these factors are typically $\simeq 1.5-3$ but can be higher 
for some models.

\begin{table}
\caption{Galactic NS-NS Coalescence Rates (Myr$^{-1}$)}
\label{table1}
\begin{tabular}{ccccl}
\tableline
\tableline
     & New& Total& Rate& Model   \\
Model& NS--NS       & NS--NS& Increase& Description \\
\tableline
A     & 3.8 & 7.5 & 2.0    & standard model described in text \\
B1    & 6.6 & 7.3 & 10     & zero kicks\\
B2    & 7.0 & 8.4 & 5.9    & single Maxwellian kicks: $\sigma=50$\,km\,s$^{-1}$\\
B3    & 5.6 & 9.5 & 2.5    & single Maxwellian kicks: $\sigma=100$\,km\,s$^{-1}$\\
D1    & 4.3 & 5.0 & 6.9    & maximum NS mass: $M_{\rm max}=2$\,M$_\odot$\\
D2    & 2.7 & 2.7 & $\gg1$ & maximum NS mass: $M_{\rm max}=1.5$\,M$_\odot$\\
E1    & 0.2 & 0.7 & 1.4    & $\alpha_{\rm CE}\times\lambda = 0.1$\\
E2    & 1.6 & 2.7 & 2.5    & $\alpha_{\rm CE}\times\lambda = 0.25$\\
E3    & 3.1 & 4.8 & 2.8    & $\alpha_{\rm CE}\times\lambda = 0.5$\\
F4    & 2.6 & 7.4 & 1.5    & mass fraction accreted: f$_{\rm a}=1.0$\\
\tableline
\end{tabular}
\end{table}

We note that the identification of the formation path for non--recycled
NS--NS binaries stems entirely from accounting for the evolution of helium
stars and for the possibility of double CE phases, both of which have
typically been ignored in previous calculations.

\section*{Conclusions}

The current theoretical estimates of NS--NS coalescence rates appear to
have a rather limited predictive power. They cover a range of values in
excess of 3 orders of magnitude. Most importantly, this range includes the
value of $\simeq 5\times 10^{-7}$\,yr$^{-1}$ required for a LIGO II
detection rate of 1 event per year. This means that at the two edges of
the range the conclusion swings from no detection to many per month, and
therefore the detection prospects of NS--NS coalescence cannot be assessed
firmly. On the other hand, empirical estimates based on the observed
sample of coalescing NS--NS systems appear to be more robust. Taking into
account recent empirical estimates and the associated uncertainties
\cite{K00}, we find the Galactic NS--NS inspiral rate in the range
$10^{-6}-5\times 10^{-4}$\,yr$^{-1}$. If we also include the independently
derived upper limit of $10^{-4}$\,yr$^{-1}$, we expect a detection rate of
$2-300$ events per year for LIGO II.

It is important to note here that another implicit assumption in
derivation of the empirical estimates is that all NS--NS binaries have at
some point in their lifetime contained a recycled pulsar with rather long
lifetimes ($\sim 10^9$\,yr). However, recent models of NS--NS formation
\cite{BK00} show that there may exist a significant NS--NS population with
neutron that never had the chance to be recycled and therefore have very
short lifetimes (by 2-3 orders of magnitude, thus preventing their
detection). For a variety of population synthesis models, the birth rate
of this separate population of coalescing NS--NS binaries is typically
comparable or higher than that of the systems with one recycled NS. The
total number of coalescing NS--NS systems could be higher by factors of at
least 50\%, and up to 10 or even higher. Such an increase has important
implications for prospects of gravitational wave detection by
ground--based interferometers. Using the recent results on the empirical
NS--NS coalescence rate \cite{K00}, we find that the {\em most optimistic}
prediction for the LIGO I detection rate could be raised to at least 1
event per 2--3 years, and the {\em most pessimistic} LIGO II detection
rate could be raised to 3--6 events per year or even higher.

Estimates of the coalescence rate of BH--NS and BH--BH systems rely solely
on our theoretical understanding of their formation. As in the case of
NS--NS binaries, the model uncertainties are significant and the ranges
extend to more than 2 orders of magnitude. However, the requirement on the
Galactic rate is less stringent for 10\,M$_\odot$ BH--BH binaries, only
$\simeq 5\times 10^{-9}$\,yr$^{-1}$. Therefore, even with the pessimistic
estimates for BH--BH coalescence rates ($\sim 10^{-7}$\,yr$^{-1}$), we
would expect at least a few to several detections per year (with LIGO II),
which is quite encouraging. We also point out that that a recent
examination of formation of close BH-BH through dynamical processes
(stellar interactions) in globular clusters leads to detection rates as
high as a few per day for LIGO II and 1 event per 2 years for LIGO I
\cite{PZ00}.

\end{document}